\documentclass[a4paper,floatfix,superscriptaddress,prx,twocolumn,showpacs,notitlepage,longbibliography]{revtex4-1}

\usepackage[colorlinks=true,citecolor=blue,urlcolor=blue]{hyperref}

\usepackage[utf8]{inputenc}
\usepackage[T1]{fontenc}
\usepackage[sc,osf]{mathpazo}
\usepackage{amsmath, amsthm, amssymb,amsfonts,mathbbol,amstext}
\usepackage{bbm}
\usepackage{mathtools}
\usepackage{comment}
\usepackage{color}
\usepackage{multirow}
\usepackage{diagbox}

\usepackage{epsfig,psfrag,amsmath,amssymb,float}
\usepackage[T1]{fontenc}
\usepackage{amsmath}
\usepackage{amssymb}

\usepackage{amscd}
\usepackage{psfrag}
\usepackage{bbm} 
\usepackage{xcolor}

\newcommand{\be}{\begin{equation}}
\newcommand{\ee}{\end{equation}}
\newcommand{\bea}{\begin{eqnarray}}
\newcommand{\eea}{\end{eqnarray}}

\begin{document}

\title{Unveiling phase transitions with machine learning}

\author{Askery Canabarro}
\affiliation{International Institute of Physics, Federal University of Rio Grande do Norte, 59078-970 Natal, Brazil}
\affiliation{Grupo de F\'isica da Mat\'eria Condensada, N\'ucleo de Ci\^encias Exatas - NCEx, Campus Arapiraca, Universidade Federal de Alagoas, 57309-005 Arapiraca-AL, Brazil
}
\author{Felipe Fernandes Fanchini}
\affiliation{Faculdade de Ci\^encias, Universidade Estadual Paulista, 17033-360 Bauru-SP, Brazil}

\author{Andr\'e Luiz Malvezzi}
\affiliation{Faculdade de Ci\^encias, Universidade Estadual Paulista, 17033-360 Bauru-SP, Brazil}

\author{Rodrigo Pereira}
\affiliation{International Institute of Physics, Federal University of Rio Grande do Norte, 59078-970 Natal, Brazil}
\affiliation{Departamento de F\'isica Te\'orica e Experimental, Federal University of Rio Grande do Norte, 59078-970 Natal, Brazil}

\author{Rafael Chaves}
\affiliation{International Institute of Physics, Federal University of Rio Grande do Norte, 59078-970 Natal, Brazil}
\affiliation{School of Science and Technology, Federal University of Rio Grande do Norte, 59078-970 Natal, Brazil}

\date{\today}
\begin{abstract}
The classification of phase transitions is a central and challenging task in condensed matter physics. Typically, it relies on the identification of order parameters and the analysis of singularities in the free energy and its derivatives. Here, we propose an alternative framework to identify quantum phase transitions, employing both unsupervised and supervised machine learning techniques. Using the axial next-nearest neighbor Ising (ANNNI) model as a benchmark, we show how unsupervised learning can detect three phases (ferromagnetic, paramagnetic, and a cluster of the antiphase with the floating phase) as well as two distinct regions within the paramagnetic phase. Employing supervised learning we show that transfer learning becomes possible: a machine trained only with nearest-neighbour interactions can learn to identify a new type of phase occurring when next-nearest-neighbour interactions are introduced. All our results rely on few and low dimensional input data (up to twelve lattice sites), thus providing a computational friendly and general framework for the study of phase transitions in many-body systems.
\end{abstract}
\maketitle

%\tableofcontents

\section{\label{sec:intro}INTRODUCTION}
%\textcolor{green}{malvezzi}

Machine learning (ML) means, essentially, computer programs that improve their performance automatically with increasing exposition to data. The algorithmic improvements over the years combined with faster and more powerful hardware \cite{aaron,highbias,breiman,altman,breiman_dt,geurts,rosenblatt,tpot,adam,ert,ensemble,rumelhart} allows now the possibility of extracting useful information out of the monumental and ever-expanding amount of data. It is the fastest-growing and most active field in a variety of research areas, ranging from computer science and statistics, to physics, chemistry, biology, medicine and social sciences \cite{Jordan}. In physics, the applications are abundant, including gravitational waves and cosmology \cite{kamdar2015,kamdar2016,kelleher,lochner,charnock,biswas,carrillo,gauci,ball,banerji, petrillo}, quantum information \cite{torlai2018neural,canabarro,iten2018discovering} and in condensed matter physics \cite{carleo2017solving,torlai2016learning,ghiringhelli2015big} most prominently in the characterization of different phases of matter and their transitions \cite{Carrasquilla,broecker2017machine,PhysRevX.7.031038,deng2017machine,huembeli2018identifying}.

Classifying phase transitions is a central topic in many-body physics and yet a very much open problem, specially in the quantum case due the curse of dimensionality of exponentially growing size of the Hilbert space of quantum systems. In some cases, phase transitions are clearly visible if the relevant local order parameters are known and one looks for non-analyticities (discontinuities or singularities) in the order parameters or in their derivatives. More generally, however, unconventional {transitions such as infinite order (e.g., Kosterlitz–Thouless) transitions}, are much harder to be identified. Typically, they appear at considerably large lattice sizes, a demanding computational task which machine learning has been proven to provide a novel approach \cite{Carrasquilla,broecker2017machine,PhysRevX.7.031038,deng2017machine,huembeli2018identifying}. For instance, neural networks can detect local and global order parameters directly from the raw state configurations \cite{Carrasquilla}. They can also be used to perform transfer learning, for example, to detect transition temperatures of a Hubbard model away from half-filling even though the machine is only trained at half-filling (average density of the lattice sites occupation) \cite{PhysRevX.7.031038}.

In this paper our aim is to unveil the phase transitions through machine learning, for the first time, in the axial next-nearest neighbor Ising (ANNNI) model \cite{Selke,Suzuki}. Its relevance stems from the fact that it is the simplest model that combines the effects of quantum fluctuations (induced by the transverse field) and competing, frustrated exchange interactions (the interaction is ferromagnetic for nearest neighbors, but antiferromagnetic for next-nearest neighbors). This combination leads to a rich ground state phase diagram which has been investigated by various analytical and numerical  approaches \cite{Villain,Allen,Rieger, Guimaraes,Beccaria,Nagy}. The ANNNI model finds application, for instance, in explaining  the magnetic order in some quasi-one-dimensional  spin ladder materials \cite{Wen}. Moreover, it has recently been used to study dynamical phase transitions \cite{Karrasch} as well as  the effects of interactions between Majorana edge modes in arrays of Kitaev chains \cite{Hassler,Milsted}.

Using the ANNNI model as a benchmark, we propose a machine learning framework employing unsupervised, supervised and transfer learning approaches. In all cases, the input data to the machine is considerably small {and simple}, the {raw} pairwise correlation functions between the spins for lattices up to $12$ sites. First, we show how unsupervised learning can detect, with great accuracy, the three main phases of the ANNNI model: ferromagnetic, paramagnetic and the clustered antiphase/floating phase. As we show, the unsupervised approach also identifies, at least qualitatively, two regions within the paramagnetic phase associated with commensurate and in-commensurate areas separated by the Peschel-Emery line \cite{Peschel}, a subtle change in the correlation functions which is hard to discern  by  conventional  methods using  only data from small chains (as we do here). Finally, we show how transfer learning becomes possible: by training the machine  with  nearest-neighbour  interactions only,  we  can  also  accurately  predict the phase transitions happening at regions including next-nearest-neighbour interactions.

The paper is organized as follows. In Sec. \ref{sec:model} we describe in details the ANNNI model. In Sec. \ref{sec:ML} we provide a succinct but comprehensive overview of the main machine learning concepts and tools we employ in this work (with more technical details presented in the Appendix). In Sec. \ref{sec:results} we present our results: in Sec. \ref{sec:dataset} we explain the data set given as input to the algorithms; in Sec. \ref{sec:unsup} we discuss the unsupervised approach followed in Sec. \ref{sec:dbscan} by an automatized manner to find the best number of clusters; in Sec. \ref{sec:sup} the supervised/transfer learning part is presented. Finally, in Sec. \ref{sec:disc} we summarize our findings and discuss their relevance.

\section{\label{sec:model}The ANNNI Model}

The axial next-nearest-neighbor Ising (ANNNI) model is defined by the Hamiltonian \cite{Selke,Suzuki}
\be
H=-J\sum_{j=1}^N\left(\sigma_j^z\sigma_{j+1}^z-\kappa \sigma_j^z\sigma_{j+2}^z+g \sigma_j^x\right).\label{ANNNI}
\ee 
Here $\sigma^{a}_j$, with $a=x,y,z$, are Pauli matrices that act on  the spin-$1/2$ degree of freedom located at site $j$  of a one-dimensional  lattice with $N$ sites and periodic boundary conditions. The coupling constant $J>0$ of the nearest-neighbor ferromagnetic exchange interaction sets the energy scale (we use $J=1$), while $\kappa$ and 
$g$ are the dimensionless coupling constants associated with the next-nearest-neighbor  interaction and the transverse magnetic field, respectively. 

The {groundstate} phase diagram of the ANNNI model exhibits four phases: ferromagnetic, antiphase, paramagnetic, and floating phase. In both the ferromagnetic  phase and the antiphase,  the $\mathbb Z_2$ spin inversion symmetry $\sigma_j^z\mapsto -\sigma_j^z$ of the model is spontaneously broken in the thermodynamic limit $N\to \infty$. However,  these two ordered phases have different order parameters. While the ferromagnetic phase is characterized by a uniform spontaneous magnetization, with one of the  ground states represented schematically by $\uparrow\uparrow\uparrow\uparrow\uparrow\uparrow\uparrow\uparrow$, the antiphase   breaks the lattice translational symmetry and has  long-range order with a  four-site periodicity in the form  $\uparrow\uparrow\downarrow\downarrow\uparrow\uparrow\downarrow\downarrow$. On the other hand, the paramagnetic phase   is    disordered    and has  a unique ground state that  can be pictured as spins pointing predominantly along the direction of the   field.  Inside the three phases described so far, the energy gap is finite and all correlation functions decay exponentially. By contrast, the floating phase   is a critical (gapless) phase with quasi-long-range order, i.e., power-law decay of   correlation functions at large distances. This phase is described by a conformal field theory with central charge $c=1$ (a Luttinger liquid  \cite{Giamarchi} with an emergent U(1) symmetry). 

The quantum phase transitions in the ANNNI model are   well understood. For $\kappa=0$, the model is integrable since it reduces to the transverse field Ising model \cite{Sachdev}. The latter is exactly solvable by mapping to noninteracting spinless fermions. Along the $\kappa=0$ line of the phase diagram, a second-order phase transition in the Ising universality class occurs at $g=1$. It separates the ferromagnetic phase at $g<1$ from the paramagnetic phase at $g>1$. Right at the critical point, the energy gap vanishes and the low-energy properties of a long chain are described by a conformal field theory with central charge $c=1/2$. 

Another simplification is obtained by setting $g=0$. In this case, the model becomes classical in the sense that it only contains $\sigma_j^z$ operators that commute with one another. For $g=0$, there is a  transition between the ferromagnetic phase at small $\kappa$ and the antiphase at large $\kappa$ that occurs exactly at $\kappa=1/2$. At this classical transition point, the ground state degeneracy grows exponentially with the system size:  any configuration that does not have three consecutive  spins pointing in the same direction is a ground state. 

For $g\neq 0$ and $\kappa\neq0$, the model is not integrable and the critical lines have to be determined numerically. In the region $0\leq\kappa\leq1/2$, the Ising transition between paramagnetic and ferromagnetic phases extends from the exactly solvable point $g=1$, $\kappa=0$ down to the macroscopically degenerate point $g=0$, $\kappa=1/2$. The latter actually becomes a multicritical point at which several critical lines meet. For fixed $\kappa>1/2$ and increasing $g>0$, one finds a second-order commensurate-incommensurate (CIC) transition \cite{Schulz}  (with dynamical exponent $z=2$) from the antiphase to the floating phase, followed by a  Berezinsky-Kosterlitz-Thouless (BKT) transition from  the floating  to the  paramagnetic phase.  

In summary, the ANNNI model has  four phases se\-parated by three quantum phase transitions. Approximate expressions for the critical lines in the phase diagram have been obtained by applying perturbation theory in the regime $\kappa<1/2$ \cite{Suzuki} or by fitting numerical  results for large chains (obtained by  density matrix renormalization group  methods \cite{Beccaria}) in the regime $\kappa>1/2$. The critical value of $g$ for the Ising transition  for $0\leq \kappa\leq 1/2$ is given approximately by \cite{Suzuki}
\be
\label{trans1}
g_{\textrm{I}}(\kappa)\approx \frac{1-\kappa}{\kappa}\left(1-\sqrt{\frac{1-3\kappa+4\kappa^2}{1-\kappa}}\right).
\ee
This approximation agrees well with the numerical estimates based on exact diagonalization for small chains \cite{Guimaraes}. The critical values of $g$ for the CIC and BKT transitions for $1/2<\kappa\lesssim 3/2$ are approximated respectively by \cite{Beccaria}
 \bea
 \label{trans2}
 g_{\textrm{CIC}}(\kappa)&\approx& 1.05\left(\kappa-0.5\right),\\
 \label{transBKT}
 g_{\textrm{BKT}}(\kappa)&\approx& 1.05\sqrt{ (\kappa-0.5 ) (\kappa-0.1)}.
 \eea
  In addition,  the paramagnetic phase is sometimes divided into two regions, distinguished by the presence of commensurate versus incommensurate oscillations in the exponentially decaying correlations. These two regions are separated by the exactly known Peschel-Emery line \cite{Peschel}, which does not correspond to a true phase  transition because the energy gap remains finite and there is no symmetry breaking  across this line. The exact expression for the Peschel-Emery line is 
\bea
\label{trans4}
g_{\textrm{PE}}(\kappa)&=& \frac{1}{4\kappa} - \kappa.
\eea
While the Ising transition  is captured correctly if one only has access to numerical results for short chains, cf. \cite{Guimaraes}, detecting the CIC and BKT transitions using  standard approaches requires computing observables for significantly longer chains \cite{Beccaria}.

\section{\label{sec:ML}Machine Learning Review}

\begin{figure}[h!]
\includegraphics*[scale=0.8]{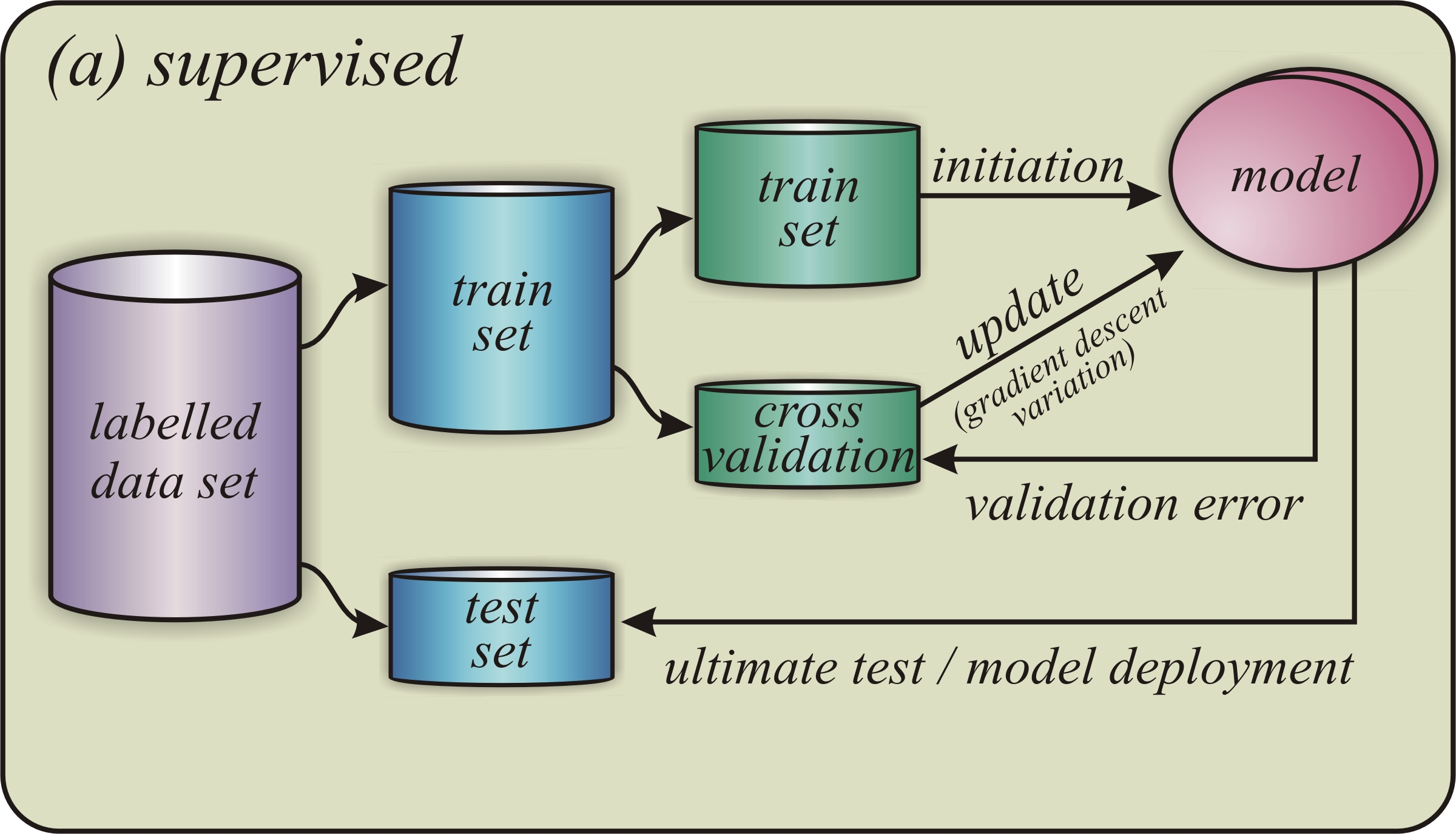}
\includegraphics*[scale=0.8]{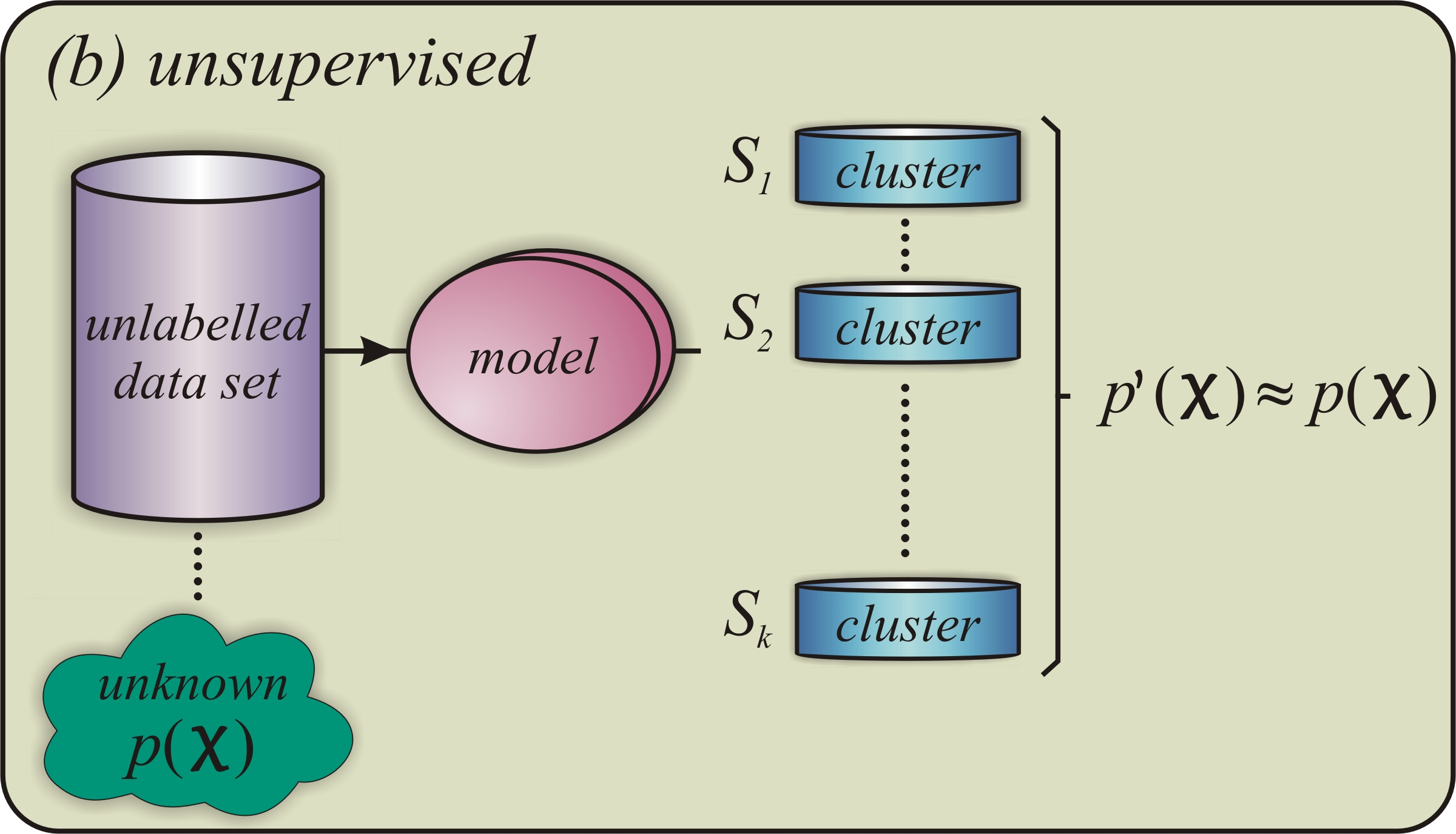}
\caption{Schematic representation of machine learning techniques. (a) Supervised learning. (b) Unsupervised learning (clustering).}
\label{fig:fig0}
\end{figure}

Machine learning is defined as algorithms that identify patterns/relations from data without being specifically programmed to. By using a variety of statistical/analytical methods, learners improve their performance $p$ in solving a task $T$ by just being exposed to experiences $E$. Heuristically speaking, machine learning occurs whenever $p(T) \propto E$, i. e. the performance in solving task $T$ enhances with increasing training data. 

The state-of-the-art of a typical ML project has four somewhat independent components: i) the data set $\mathbf{X}$, ii) a model $m(\mathbf{w})$, iii) a cost function $J(\mathbf{X};m(\mathbf{w}))$ and iv) an optimization procedure. Our aim is to find the best model parameters, $\mathbf{w}$, which minimizes the cost function for the given data set. This optimization procedure is, generally, numerical and uses variations of the well known gradient descent algorithm. In this manner, by combining distinct ingredients for each component in this recipe, we end up with a myriad of possible machine learning pipelines \cite{aaron,highbias}. 

In machine learning research, there are two main approaches named as supervised and unsupervised learning, related to the experience passed to the learner. The central difference between them is that supervised learning is performed with prior knowledge of the correct output values for a subset of given inputs. In this case, the objective is to find a function that best approximates the relationship between input and output for the data set. In turn, unsupervised learning, does not have labeled outputs. Its goal is to infer the underlying structure in the data set, in other words, discover hidden patterns in the data, see Fig. \ref{fig:fig0} for a pictorial distinction. In this work, in order to propose a general framework, we used both approaches in a complementary manner.

Using the supervised learning as a prototype, one can depict the general lines of a ML project. We first split the data set into two non-intersecting sets: $\mathbf{X}_{\text{train}}$ and $\mathbf{X}_{\text{test}}$, named training and test sets, respectively. Typically, the test set corresponds to $10 \%-30 \%$ of the data set. Then, we minimize the cost function with the training set, producing the model, $m(\mathbf{w}^*)$, where $\mathbf{w}^* = \text{argmin}_w \{J(\mathbf{X}_{\text{train}};m(\mathbf{w}))\}$. We evaluate the cost function of this model in the test set in order to measure its performance with out-of-sample data. In the end, we seek a model that performs adequately in both sets, i. e., a model that generalizes well. In other words, the model works fine with the data we already have as well as with any future data. This quality, called generalization, is at the core of the difference between the machine learning approach and a simple optimization solution using the whole data set.   

The performance of the model is made by evaluating the cost function in the test set. This is generally done by computing the mean square error (MSE) between the prediction made by the model and the known answer (target), $\epsilon_{\text{test}} = \langle J(\mathbf{X}_{\text{test}};m(\mathbf{w}^*))\rangle$. In a ML project, we are dealing, generally, with complex systems for which we have \textit{a priori} no plausible assumption about the underlying  mathematical model. Therefore, it is common to test various types of models ($m_1,m_2, ...$) and compare their performance on the test set to decide which is the most suitable one. In fact, it is even possible to combine them (manually or automatically) in order to achieve better results by reducing bias and variance \cite{aaron,highbias,canabarro}.

One should be careful with what is generally called overfitting, that is, some models may present small values for $\epsilon_{\text{train}}$, but $\epsilon_{\text{test}} \gg \epsilon_{\text{train}}$. It happens because some mo\-dels (often very complex) can deal well with data we already have, however produce large error with unobserved data. Overfitting is a key issue in machine learning and various methods have been developed to reduce the test error, often causing an increase in the training error, but reducing the generalization error as a whole. Making ensemble of multiple models is one of such techniques and as has been already successfully demonstrated \cite{aaron, canabarro}. On the opposite trend, we might also have underfitting, often happening with very simple models where $\epsilon_{\text{test}} \sim \epsilon_{\text{train}}$ are both large. Although extremely important, the discussion about the bias-variance trade-off is left to the good review in Refs. \cite{aaron,highbias}.

Overall, the success of a ML project depends on the quality/quantity of available data and also our prior knowledge about the underlying mechanisms of the system. In the Appendix, we provide a brief but intuitive description of all machine learning steps involved in our work. These include the tasks (classification and clustering), the experiences (supervised and unsupervised learning), the machine learning algorithms (multi-layer perceptron, random forest, and so on) and also the performance measures. For additional reading and more profound and/or picturesque discussions we refer to \cite{aaron,highbias} and as well as to the Appendix.

\section{\label{sec:results}Machine Learning phase transitions in the ANNNI model}

\subsection{\label{sec:dataset}Our Data Set}
We use the pairwise correlations among all spins in the lattice as the data set to design our models. Thus, the set of observables used is given by $\left\{ \langle \sigma^{x}_{i}\sigma^{x}_{j} \rangle,\langle \sigma^{y}_{i}\sigma^{y}_{j } \rangle, \langle \sigma^{z}_{i}\sigma^{z}_{j} \rangle \right\}$ with, $j>i$ and $i=[1,N-1]$ where $N$ is the number of spins/qubits in the lattice and $\langle \sigma^{x}_{i}\sigma^{x}_{j} \rangle =\langle\lambda_{0}|\sigma_{i}^{x}\sigma_{j}^{x}|\lambda_{0}\rangle$ is the expectation value of the spin correlation for the Hamiltonian ground state $\vert \lambda_{0}\rangle$ (analogously to {$\langle \sigma^{y}_{i}\sigma^{y}_{j} \rangle$ and $\langle \sigma^{z}_{i}\sigma^{z}_{j} \rangle$}).
It is easy to see that the number of features is given by $3\sum_{k=1}^{N-1}k$ since that for $8$, $10$, and $12$ sites we have $84$, $135$, and $198$ features respectively.

\subsection{\label{sec:unsup}Unsupervised Approach}

Unsupervised learning is the branch of machine learning dealing with data that has not been labeled, classified or categorized. Simply from the features (the components) of the input data, represented by a vector $\mathbf{X}$, one can extract useful properties about it. Quite generally, we seek for the entire probability distribution $p(\mathbf{X})$ that generated and generalizes the data set. Clustering the data into groups of similar or related examples is a common task in an unsupervised project. Self-labeling is another crucial application of unsupervised learning, opening the possibility of combining unsupervised and supervised algorithms to speed up and/or improve the learning process, an approach known as semi-supervised learning (SSL). As we shall demonstrate, using the ANNNI model as a benchmark, unsupervised learning offers a valuable framework to analyze complex phase diagrams even in situations where only few and low dimensional input data  is available.

Briefly describing, the algorithm is used to partition $n$ samples into $K$ clusters, fixed \textit{a priori}. In this manner, $K$ centroids are defined, one for each cluster. The next step is to associate each point of the data set to the nearest centroid. Having all points associated to a centroid, we update the centroids to the barycenters of the clusters. So, the $K$ centroids change their position step by step until no more changes are done, i.e. until a convergence is achieved. In other words, centroids do not move more {within} a predetermined threshold. See Appendix for more details.

\begin{figure}[t!]
\includegraphics*[scale=0.6]{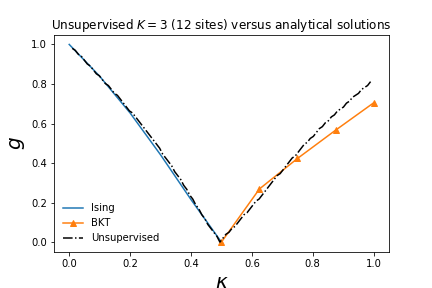}
\caption{
Comparison among the approximate solutions \eqref{trans1} and \eqref{transBKT} and the unsupervised learning trained  $N=12$ sites in the lattice. The Ising transition \eqref{trans1} is almost perfectly reproduced. The machine results for the BKT transition  \eqref{transBKT} shows a smaller accuracy; nonetheless, it is qualitatively accurate.}
\label{fig:fig1}
\end{figure}

For distinct pairs of the coupling parameters ($g;\kappa$) of the Hamiltonian \eqref{ANNNI}, we explicitly compute all the pairwise spin correlations described in section ~\ref{sec:dataset}. Since the computation of correlations is computationally very expensive, the coupling parameters were varied with step size of $10^{-2}$ in the range $\kappa, g \in [0,1]$. So, in total, we are training the learner with a modest number of 10000 examples. Equipped with that, we investigate the capacity of an unsupervised algorithm to retrieve the phase diagram. In Fig.~\ref{fig:fig1}, we show the phase diagram produced by using $k$-means algorithm \cite{kmeans} and focusing on a lattice size $N=12$. Providing $K=3$ to the learner (assuming three phases), the algorithm returns the best clustering based on the similarities it could find in the features. Strikingly, given the few data points and the relatively small size of spin chain used to generated the data, it finds three well distinct clusters in very good accordance with the three main phases of the ANNNI model. 
Indeed, since we imposed to the method the gathering in only three groups, the K-means algorithm detects the ferromagnetic phase, the paramagnetic phase, and a third one which clustered the floating phase with the antiphase. It is quite surprising because the boundary between the paramagnetic and the floating phases is the BKT transition, therefore it detects a transition which is notoriously hard to pinpoint, as the correlation length diverges exponentially at the critical point \cite{koster1}. Moreover, the unsupervised approach almost perfectly recovers the curve corresponding to the ferromagnetic-paramagnetic transition and its analytical critical value of $g_{\mathrm{crit}}=1$ (with $\kappa = 0$). As well, it gives very accurate quantitative predictions for the analytical tricritical value $\kappa_{\mathrm{crit}}=1/2$, at which the transition between the ferromagnetic, paramagnetic and antiphase regions happens.  

\begin{figure}[t!]
\includegraphics[scale=0.6]{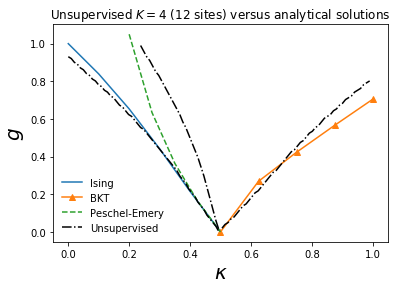}
\caption{Comparison among the approximate solutions \eqref{trans1}, \eqref{transBKT} and \eqref{trans4} with the unsupervised results for $K=4$. As one can see, at least qualitatively, the machine is finding a new region associated with the Emery-Peschel line \eqref{trans4}.} 
\label{fig:fig1b}
\end{figure}

We have also tested the unsupervised prediction by setting $K=4$, that is, assuming four phases for the ANNNI model as described in Section \ref{sec:model}. The result is shown in Fig. \eqref{fig:fig1b}.
As we can see, the algorithm does not separate the floating phase and the antiphase but, instead, the new critical line that appears for $K=4$ divides the paramagnetic phase into two regions which, at least qualitatively, can be identified with the commensurate and incommensurate  regions separated by the Peschel-Emery line  \eqref{trans4}. This result is remarkable because it tells us that machine learning approach manages to detect a subtle change in the correlation functions which is hard to discern by conventional methods using only data for small chains, up to $N=12$. On the other hand, recall that such change in the correlation function does not correspond to a phase transition in the strict sense.

The results presented in this section indicate that unsupervised learning approach is a good candidate when one
knows in advance a good estimate for the number of phases $K$, as it is requested upfront for various unsupervised algorithms. In the next sections, we show how supervised learning can be used as a validation step for the unsupervised results, also providing surprisingly accurate results. However, we first address the task of how we can use a complementary unsupervised approach to cope with the limitation mentioned above, that is, when one has no \textit{a priori} knowledge of a reasonable number of existing phases.

\subsection{\label{sec:dbscan}Density-based clustering}

To make our framework applicable to cases in which we have no guess of how many phases we {can} possibly expect, we propose to use a density-based (DB) clustering technique \cite{dbscan2,dbscan} to estimate an initial number of clusters. Density clustering makes the intuitive assumption that clusters are defined by regions of space with higher density of data points, meaning that outliers are expected to form regions of low density. The main input in such kind of algorithms is the critical distance $\epsilon$ above which a point is taken as an outlier of a given cluster. In fact, it corresponds to the maximum distance between two samples for them to be labelled as in the same neighborhood. One of its most relevant output is the estimated number of distinct labels, that is, the number of clusters/phases. Therefore, it can be taken as a complementary technique for the use of the $k$-means or any other unsupervised approach which requires one to specify the number of clusters expected in the data.

DBSCAN is one of the most common DB clustering algorithms and is known to deal well with data which contains clusters of similar density \cite{dbscan}, making it suitable for our case. We use DBSCAN to retrieve the number of clusters we should input in the unsupervised KNN algorithm, thus assuming no prior knowledge of how many phases one expects. For that, we feed the DBSCAN with a critical distance $\epsilon$ of the order of the step size used to span the training data, i. e., $\epsilon = 10^{-2}$. As a result, the algorithm returned 3 clusters as the optimal solution, thus coinciding with the three main phases present in the ANNNI model, precisely those that one can expect to recognize at the small lattice size we have employed. It also coincides with the Elbow curve for estimating the optimal number of clusters, see Appendix for details.

\subsection{\label{sec:sup}Supervised Approach}

In spite of the clear success of unsupervised ML in identifying the phases in the ANNNI model, a natural question arises. How can we trust the machine predictions in the absence of a more {explicit} knowledge about the Hamiltonian under scrutiny? {Could} partial knowledge help in validating the ML results?
Typically, limiting cases of a Hamiltonian of interest
are simpler and thus more likely to have a known solution. That is precisely the case of the ANNNI model, which for $\kappa=0$ is fully understood, in particular the fact that at $g=1$ there is a phase transition between the ferromagnetic and paramagnetic phases. Can a machine trained with such limited knowledge ($\kappa=0$) make any meaningful predictions to the more general model ($\kappa \geq 0$)? 
The best one can hope for in this situation is that the unsupervised and supervised approaches point out similar solutions, a cross validation enhancing our confidence. In the following we show that this is indeed possible by investigating supervised learning algorithms as a complementary approach to the unsupervised framework introduced above. 

In supervised machine learning, the learner experiences a data set of features $\mathbf{X}$ and also the target or label vector $\mathbf{y}$, provided by a "teacher", hence the term "supervised". In other words, the learner is presented with example inputs and their known outputs and the aim is to create a general rule that maps inputs to outputs, by generally estimating the conditional probability $p(\mathbf{y} | \mathbf{X})$. One of the main differences between a ML algorithm and a canonical algorithm is that in the second we provide inputs and rules and receive answers and in the first we insert inputs and answers and retrieve rules (see Appendix for more details). 

Our aim is to understand whether transfer learning is possible (training with $\kappa=0$ to predict at regions where $\kappa \geq 0$). Both the unsupervised approach as well as the analytical solution to $\kappa=0$, point out that a transition occurs at $g \approx  1$. With this information, we train the supervised algorithms with $g$ ranging in the interval $[0.5,1.5]$. Given that we don't have to vary over $\kappa$, we reduce the step size (in comparison with the unsupervised approach) to $10^{-3}$, generating an evenly distributed training data with equal number of  samples, 500 in each phase (ferromagnetic for $g < 1$ and paramagnetic for $g> 1$ ). The main drawback of this supervised approach is that it always performs binary classification, known as one-vs-all classification.  For instance, for handing writings digits it is similar to the case in which a learner can simply identify whether or not the number 5 has been written.

\begin{figure}[t!]
\begin{center}
\includegraphics*[scale=0.6]{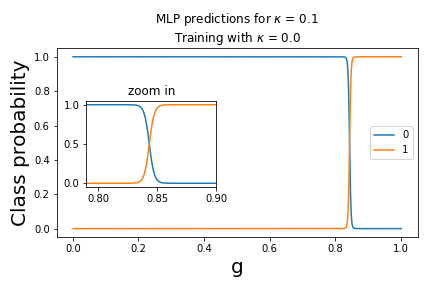}
\end{center}
\caption{Detecting the critical transverse magnetic field coupling parameter $g$ at which a phase transition occurs. The machine was trained at $\kappa=0$ and asked to predict where the transition happens at $\kappa=0.1$, by considering where the machine is most uncertain, that is, when the probabilities $p_1=p_2=1/2$. Here the ferromagnetic (paramagnetic) phase is labeled as $0$ $(1)$.}
\label{fig:fig3}
\end{figure}

\begin{figure}[t!]
\begin{center}
\includegraphics*[scale=0.6]{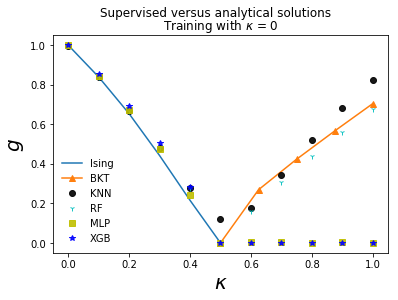}
\end{center}
\caption{
Phase diagrams produced with diverse ML algorithms when trained only with $\kappa =0$: KNN (black circles), Random Forest (cyan down stars), Multilayer Perceptron (yellow squares) and  Extreme Gradient Boosting (blue diagonal stars) and two different analytical solutions (Ising (solid blue) and BKT (solid orange triangles). All different methods recover the ferro/paramagnetic transition very well while the transition between the paragmanetic and the BKT are only recovered by the KNN and RF methods (see main text for more details).}
\label{fig:fig2}
\end{figure}

Motivated by the sound results in its unsupervised version, we first tried the KNN algorithm (vaguely related to the $k$-means method) as well as different methods such as the multilayer perceptron (MLP, a deep learning algorithm), random forest (RF) and extreme gradient boosting (XGB). Once the model is trained, we use the same data set used in section ~\ref{sec:unsup} to predict the corresponding phases. Actually, for a given instance $X^{'}$, the trained model, $m$, returns $m(X^{'})= (p_1,p_2)$, where $(p_1,p_2)$ is a normalized probability vector and the assigned phase corresponds to the component with largest value. To determine when we are facing a transition, we plot both the probability components and check when they cross, as shown in Fig.~\ref{fig:fig3}. As can be seen in Fig.~\ref{fig:fig2}, the different ML methods successfully recover the left part ($\kappa < 0.5$) of the phase diagram, exactly corresponding to the ferromagnetic/paramagnetic transition over which the machine has be trained for the case $\kappa=0$. However, what happens as we approach the tricritical point at $\kappa=0.5$ at which new phases (the antiphase and the floating phase) appears? As can be seen in Fig.~\ref{fig:fig2}, near this point the predictions of the different methods start to differ.

To understand what is going on, we highlight that since the machine can only give one out of two answers (the ones it has been trained for), the best it can do is to identify {the clustered antiphase/floating phase (here labeled as phase '2')} with either the ferromagnetic (phase '0') or the paramagnetic (phase '1') cases. Since the models were never trained for the antiphase, it is a good test to check the learner's ability to classify a new phase as an outlier. 
For $\kappa \geq 0.5$, we are in a region where only phases '0' and '2' are present. So, the best the machine model can do is to output '0' fs the phase is indeed '0' and '1' otherwise. As we already remarked, it is a drawback of the supervised approach in comparison to the unsupervised one, but it is still useful to validate the emergence of a new phase, as suggested by the unsupervised technique. In this sense, the KNN and RF methods perform quite well. 
As seen in Fig.~\ref{fig:fig2} the transition between the paramagnetic and the clustered antiphase/floating is qualitatively recovered even though the machine has never been exposed to these phases before. %\textcolor{blue}{EU REMOVI A FIGURA E ESTA PARTE A SEGUIR DO TEXTO. SE FOR PARA COLOCAR A FIG. PRECISA REESCREVER ESTA PARTE. AO MEU VER FICA CLARO E SIMPLE ASSIM.} \sout{The transition point is illustrated in Fig.~\ref{fig:fig4} (Bottom) considering $\kappa=0.9$.  In turn, the MLP and XGB methods also perceive a change at tricritical point $\kappa=0.5$. However, their performance is worst, the antiphase being recognized as the paramagnetic one as can be seen in Fig.~\ref{fig:fig2} and Fig.~\ref{fig:fig4} (Top).} 

%%%%%%%%%%%%%%%%%%%%%%%%%%%%%%%%%%%%%%%%%%%%%
%EU REMOVERIA ESTA FIGURA
%%%%%%%%%%%%%%%%%%%%%%%%%%%%%%%%%%%%%%%%%%%%%
%\begin{figure}[t!]
%\begin{center}
%\includegraphics[scale=0.6]{fig5_vf.png}
%\includegraphics*[scale=0.6]{trans_k09_train_k0_KNN_veer2.png}
%\end{center}
%\caption{Detecting the critical transverse magnetic field coupling parameter $g$ with the  components of the probability vector for a binary classification problem for $\kappa = 0.9$ trained with $\kappa = 0$ delivered by a MLP (Top) and supervised KNN learner (Bottom). \textcolor{green}{O texto na parte de baixo diz treinamento com $\kappa = 0.6$ mas seria $\kappa = 0$ certo? Na versão de submissão creio que todos esses textos acima das figuras tem que ser retirados. Só os captions ficam.}}
%\label{fig:fig4}
%\end{figure}
%%%%%%%%%%%%%%%%%%%%%%%%%%%%%%%%%%%%%%%%%%%%%

In Table 1 we present the average {$\ell_1$-norm} for the distinct algorithms taking as benchmark the approximate analytical solutions for the three main phases given by Eqs. \eqref{trans1} and \eqref{transBKT}. One can observe that the two best approaches are the supervised RF and the unsupervised KNN trained with 12 sites, which reinforces our framework of using the unsupervised and supervised methodologies complementarily. It is worth mentioning that the better performance of the supervised approach is related to the fact that we provided more training data, accounting for a more precise transition, as the step size over $g$ is reduced. %\textcolor{red}{[Explain in more details how this L1 norm was computed and include the results for when $\kappa=0.6$ is used as well.]}

\begin{table}[!t]
\caption{Performance (average $\ell_1$-norm with relation to the analytical approximations given by Eqs. 2 and 3) computed for the three main phases and different ML approaches. See Appendix for details. Two best ones in boldface.} 
\label{general_analysis}
\setlength\tabcolsep{0pt} 
\footnotesize\centering
\smallskip 
\begin{tabular*}{\columnwidth}{@{\extracolsep{\fill}}lcc}
\hline
\diagbox[width=8em]{Technique}& average $\ell_1$-norm \\ 
 \hline
 \hline
RF (supervised) & \textbf{0.03375(9)} \\
KNN (supervised) & 0.07461(4) \\
MLP (supervised) & 0.18571(4) \\
XGB (supervised) & 0.19507(7) \\
KNN (unsupervised - 12 sites) & \textbf{0.03474(4)} \\
KNN (unsupervised - 8 sites)  & 0.07904(1) \\
KNN (unsupervised - 10 sites) & 0.16350(2) \\
\end{tabular*}
\end{table}

\section{Discussion}
\label{sec:disc}
In this paper we have proposed a machine learning framework to analyze the phase diagram and phase transitions of quantum Hamiltonians. Taking the ANNNI model as a benchmark, we have shown how the combination of a variety of ML methods can characterize, both qualitatively and quantitatively, the complex phases appearing in this Hamiltonian. First, we demonstrated how the unsupervised approach can recover accurate predictions about the 3 main phases {(ferromagnetic, paramagnetic and the clustered antiphase/floating)} of the ANNNI model. It can also recover qualitatively different regions within the paramagnetic phase as described by the Emery-Peschel line, even though there is no true phase transition in the thermodynamic limit. This is remarkable, given that typically this transition needs large lattice sizes to be evidenced. Here, however, we achieve that using comparatively very small lattice sizes (up to 12 spins). Finally, we have also considered supervised/transfer learning and showed how transfer learning becomes possible: by training the machine at $\kappa=0$ ($\kappa$ representing the next-neighbour interaction strength) we can also accurately {determine} the phase transitions happening at $\kappa \geq 0$. %\textcolor{red}
{Surprisingly, the machine trained to distinguish the ferromagnetic and paramagnetic phase is also able to identify the BKT transition, which is
notoriously hard to pinpoint, since the correlation length
diverges exponentially at the critical point. Moreover, we note that the machine is not simply testing the order parameter because it is able to distinguish between the ferromagnetic phase and the antiphase which are both ordered but with a different order parameter. Indeed, the machine is identifying a new pattern in the data.}

%\textcolor{blue}
{Taken together, these results suggest that the ML investigation should start with the unsupervised techniques (DBSCAN + Clustering), as it is a great initial exploratory entry point in situations where it is either impossible or impractical for a human or an ordinary algorithm to propose trends and/or have insights with the raw data. After some hypotheses are collected with the unsupervised approach, one should now proceed to get more data and apply supervised techniques as to consolidate the unsupervised outputs. See Appendix for more discussions.}

Overall, we see that both the unsupervised and the various supervised machine learning predictions are in very good agreement. Not only do they recover   similar critical lines but also discover the multicritical point of the phase diagram. Clearly, we can only say that because the precise results are known for the ANNNI model. The general problem of mapping out phase diagrams of quantum many-body systems is still very much open, even more so with the discovery of topological phases. The methods investigated here may contribute to advancing this field and hope to motivate the application of this framework to Hamiltonians where the different phases have not yet been completely sorted out. If as it happens for the ANNNI model, if all these multitude of evidence obtained by considerably different ML methods point out similar predictions, this arguably give us good confidence about the correctness of the results. The predictions given by the ML approach we propose here can be seen as guide, an initial educated guess of regions in the space of physical parameters where phase transitions could be happening. 

\section*{Acknowledgements}
The authors acknowledge the Brazilian ministries MEC and MCTIC, funding agency CNPq (AC's Universal grant No. $423713/2016-7$, RC's grants No. $307172/2017-1$ and $No 406574/2018-9$ and INCT-IQ) and UFAL (AC's paid license for scientific cooperation at UFRN). This work was supported by the Serrapilheira Institute (grant number Serra-1708-15763) and the John Templeton Foundation via the grant Q-CAUSAl No 61084.

\bibliography{refs}

\section*{Appendix}

\subsection{The experience E: supervised and unsupervised learning}

Within machine learning research, there are two major methods termed as supervised and unsupervised learning. In a nutshell, the key difference between them is that supervised learning is performed using a firsthand information, i. e., one has ("the teacher", hence the term "supervised") prior knowledge of the output values for all and every input samples. Here, the objective is to find/learn a function that best ciphers the relationship between input and output. In contrast, unsupervised learning deals with unlabeled outputs and its goal is to infer the natural structure present within the data set.

In a supervised learning project, the learner experiences data points of features $\mathbf{X}$ and also the corresponding target or label vector $\mathbf{y}$, aiming to create a universal rule that maps inputs to outputs, by generally estimating the conditional probability $p(\mathbf{y} | \mathbf{X})$. Typical supervised learning tasks are: (i) classification, when one seeks to map input to discrete output (labels), or (ii) regression, when one wants to map input to a continuous output set. Common supervised learning algorithms include: Logistic Regression, Naive Bayes, Support Vector Machines, Decision Tree, Extreme Gradient Boosting (XGB), Random Forests and Artificial Neural Networks. Later, we provide more details of the ones used in this work. 

In unsupervised learning, as there is no teacher to label, classify or categorize the input data, the learner must grasp by himself how to deal with the data. Simply from the features (the components) of the input data, represented by a vector $\mathbf{X}$, one can extract useful properties about it. Quite generally, we seek for the entire probability distribution $p(\mathbf{X})$ that generated and generalizes the data set, as we have pointed out in the main text. The most usual tasks of an unsupervised learning project are: (i) clustering, (ii) representation learning \cite{RL} and dimension reduction, and (iii) density estimation. Although we focus on clustering in this work, in all of these tasks, we aim to learn the innate structure of our unlabelled data. Some popular algorithms are: K-means for clustering; Principal Component Analysis (PCA) and autoencoders for representation learning and dimension reduction; and Kernel Density Estimation for density estimation. Since no labels are given, there is no specific way to compare model performance in most unsupervised learning methods. In fact, it is even hard to define technically Unsupervised Learning, check \cite{aaron,highbias} for deeper discussions.

As briefly mentioned in the main text, unsupervised learning is very useful in exploratory analysis and integrated many ML pipelines once it can automatically identify structure in data. For example, given the task of labelling or inserting caption for the tantamount of images available on the world web, unsupervised methods would be (and in fact is) a great starting point for this task, as those used by Google's Conceptual Captions Team. In situations like this, it is impossible or at least impractical for a human or even a regular algorithm to perform such task.

Machine learning is a method used to build complex models to make predictions in problems difficult to solve with conventional programs. It may as well shed new light on how intelligence works. However, it is worth mentioning that this is not, in general, the main purpose behind machine learning techniques given that by "learning" it is usually meant the skill to perform the task better and better, not necessarily how it is learning the task itself. For instance, in a binary classification task, e. g., "cat vs dog", the program does not learn what a dog is in the same sense as human does, but it learns how to differentiate it from a cat. In this manner, it only learns the following truisms: (i) a dog is a dog and (ii) a dog is not a cat (and similar statements for what a cat is). Also, it is important to emphasize that a highly specialized learning process does not imply, necessarily, a better instruction. For example, a more flexible process may, in fact, be more efficient to achieve an efficient transfer learning.

\subsection{Tasks: classification and clustering}

A machine learning task is specified by how the learner process a given set of input data  $\mathbf{X} \in \mathbb{R}^{n}$. Some usual ML tasks are: classification, regression, clustering, transcription, machine translation, anomaly detection, and so on. Here, we quickly describe the two kinds of tasks we use in this work: classification and clustering.

\subsubsection{Classification}

Here, the program must designate in which class an input instance should be classified, given $K$ possible categories. The algorithm retrieves a function $f: \mathbb{R}^{n} \to \{1, ..., k\}$, where $K$ is a finite and commonly pre-set integer number. So, for a given input vector $\mathbf{X}$, the model returns a target vector $\mathbf{y} = f(\mathbf{X})$. Typically, $f$ returns a normalized probability distribution over the $K$ classes and the proposed class is that with highest probability. Referring to the main text, this is the case for all the supervised algorithms.       

\subsubsection{Clustering}

Clustering is one of the most important unsupervised learning problem. As the other problems of this approach, it aims to identify a structure in the unlabelled data set. A loose definition of clustering could be “the process of organizing objects into groups whose members are similar in some way”. A cluster is therefore a collection of objects which are “similar” between them and are “dissimilar” to the objects belonging to other clusters. Later in this Appendix, we provide more detail about $k$-means as well as density-based clustering. However,  for a more technical definition, please check Ref. \cite{dbscan2}. For a formal mathematical description, we refer the reader to Ref. \cite{macqueen}.

\subsection{The performance p}

One important aspects of supervised learning which makes it different from an optimization algorithm is \textit{generalization}, as mentioned in section \ref{sec:ML}. It means that we project the algorithm to perform well both on seen (train set) and unseen (test set) inputs. This is accomplished by measuring the performance on both the sets. The test set normally corresponds to $10 \%-30 \%$ of the available data.  

For a binary classification tasks (as the ones we dealt with in this work), one practical measure is the accuracy score which is the proportion of correct predictions produced by the algorithm. For the classification task in this work, we were satisfied with a specific model, i. e. the model was taken to be optimal, when both the train and test sets accuracy score were higher than $99.5 \%$. It is worth mentioning that we used these well-trained models for $\kappa = 0$ when we seek to achieve an efficient transfer learning. In other words, when we tried to used these models for  $\kappa > 0$

As already addressed in the main text, the optimal model is found by minimizing a cost function. However, the ideal cost function varies from project to project and from algorithm to algorithm. For all supervised techniques, we used the default cost function of the Python scikit-learn package for the implementation of a given algorithm \cite{scikit}. Therefore, we present the cost functions case by case in the next subsection. However, to present the performance in Table \ref{general_analysis}, we use the L$_1$ error due to its straightforward interpretation for model comparison.   

\subsection{Algorithms}

All algorithms used in this work are native to the Python Scikit-Learn package, see Ref. \cite{scikit} for more details. Here, we provide a description of the used methods, highlighting the parameters and hyper-parameters used to calibrate the models and consequently generate the figures.  

\subsubsection{$k$-means clustering} 

The $k$-means is an unsupervised learning algorithm. It searches for $k$ clusters within an unlabeled multidimensional data set, where the centers of the clusters are taken to be the arithmetic mean of all the points belonging to the respective cluster, hence the term $k$-means. Clustering stands for finding groups of similar objects while keeping dissimilar objects in different groups \cite{dbscan2}. When convergence is reached, each point is closer to its own cluster center than to other cluster centers. It can be seen as a vector quantization (discrete output), once its final product it to assert labels to the entire data set. 

\begin{figure}[h!]
\includegraphics*[scale=1]{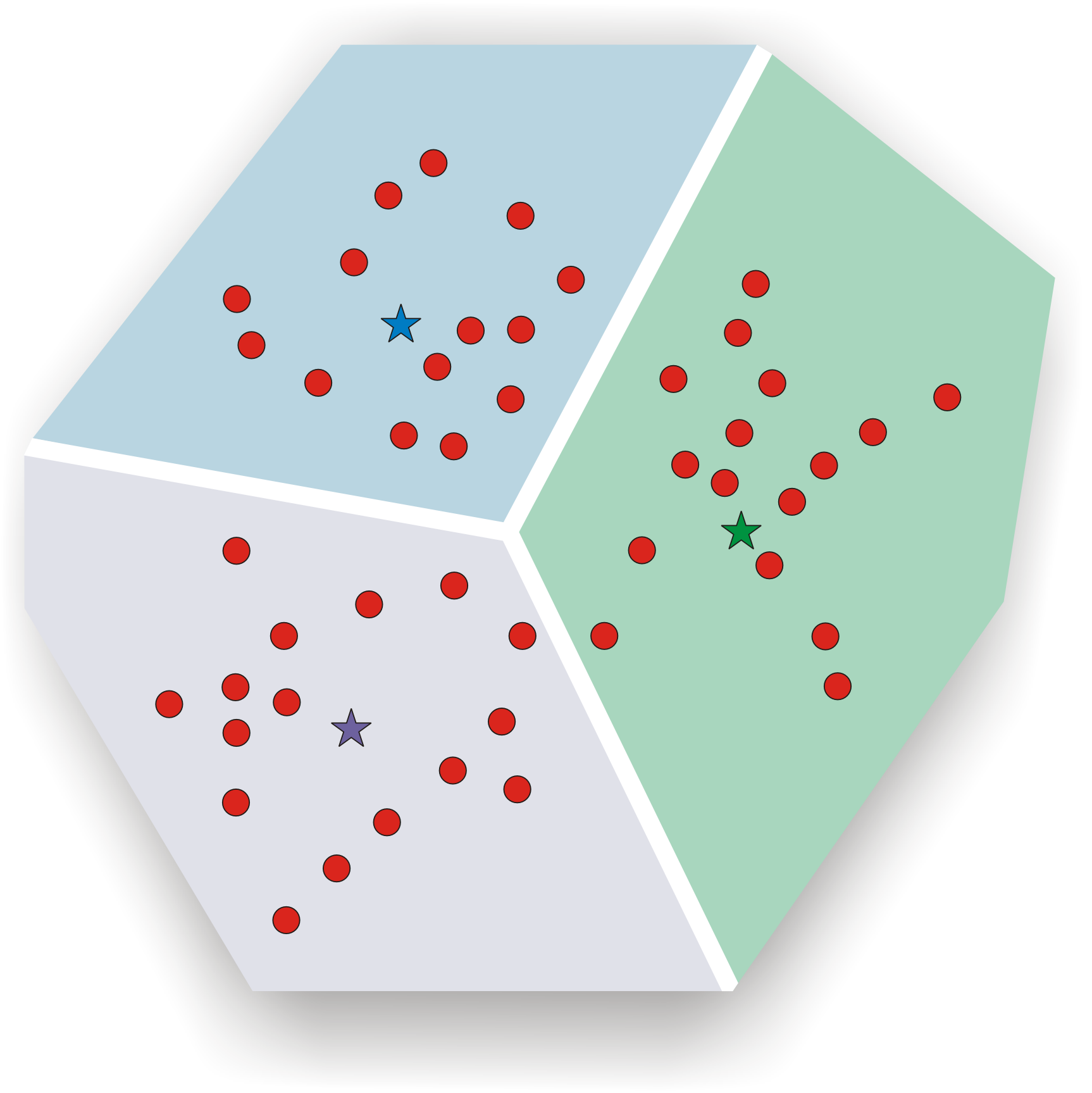}
\caption{Schematic representation of the $k$-means clustering.}
\label{fig:kmeans}
\end{figure}

The data set is, generally, a set of $d$-dimensional real-valued points sampled from some unknown probability distribution and the dissimilarity function is often the Euclidean distance. $k$-means clustering aims to partition $n$ observations into $k$ clusters in which each observation belongs to the cluster with the nearest mean, serving as a prototype of the cluster and providing a meaningful grouping of the data set. This results in a partitioning of the data space into Voronoi cells, as schematically represented in Fig. \ref{fig:kmeans}.

Given a set of $n$ observations $(X_1, X_2, ..., X_n)$, where $X_i \in \mathbb{R}^d$, $k$-means clustering seeks to partition the $n$ observations into $k (k \le n)$ sets $S = {S_1, S_2, ..., S_k}$ so as to minimize the within-cluster sum of squares, which happens to be the variance. Formally, the objective is to find:

\begin{equation}
{\displaystyle {\underset {\mathbf {S} }{\operatorname {arg\,min} }}\sum _{i=1}^{k}\sum _{\mathbf {x} \in S_{i}}\left\|\mathbf {x} -{\boldsymbol {\mu }}_{i}\right\|^{2}={\underset {\mathbf {S} }{\operatorname {arg\,min} }}\sum _{i=1}^{k}|S_{i}|\operatorname {Var} S_{i},}    
\end{equation}
where $\boldsymbol{\mu}_j$ is the mean of the points in $S_i$. This is equivalent to minimizing the pairwise squared deviations of points in the same cluster:

\begin{equation}
{\displaystyle {\underset {\mathbf {S} }{\operatorname {arg\,min} }}\sum _{i=1}^{k}\,{\frac {1}{2|S_{i}|}}\,\sum _{\mathbf {x} ,\mathbf {y} \in S_{i}}\left\|\mathbf {x} -\mathbf {y} \right\|^{2}}. 
\end{equation}

The algorithm runs as follows: (1) (randomly) initialize the $k$ clusters centroids $\boldsymbol{\mu}_j$; (2) [repeat until convergence] a) for every observation $X_i$ assign it to the nearest cluster; b) update the cluster centroid. Check \cite{macqueen} for detailed mathematical demonstration of asymptotic behavior and convergence. Intuitively, one seeks to minimize the within-group dissimilarity and maximize the between-group dissimilarity for a predefined number of clusters $k$ and dissimilarity function. From a statistical perspective, this approach can be used to retrieve $k$ probability densities $p_i$. Assuming that they belong to the same parametric family (Gaussian, Cauchy, and so on), one can take the unknown distribution $p$ as a mixture of the $k$ distribution $p_i$, as depicted in figure \ref{fig:fig1} (b).

We have used the library \textit{sklearn.cluster.KMeans} which uses some new initialization technique for speeding up the convergence time. The main input parameter is the number of clusters. Overall, $k$-means is a very fast algorithm. Check the library for the details and the default parameters \cite{scikit}.

\subsubsection{Density-based (DB) clustering} 

In density-based clustering methods the clusters are taken to be high-density regions of the probability distribution, giving reason for its name \cite{dbscan2}. Here, the number of clusters is not required as an input parameter. Also, the unknown probability function is not considered a mixture of the probability functions of the clusters. It is said to be a nonparametric approach. Heuristically, DB clustering corresponds to partitioning group of high density points (core points) separated by contiguous region of low density points (outliers).

\begin{figure}[t]
\includegraphics*[scale=0.6]{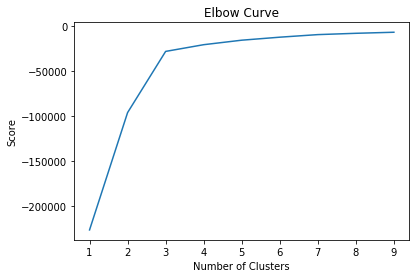}
\caption{Elbow curve for the estimation of the best $k$ for $k$-means clustering method.}
\label{fig:elbow}
\end{figure}

In this work, we used the Density-Based Spatial Clustering (DBSCAN) using the scikit-learn package \textit{sklearn.cluster.DBSCAN}. The algorithm is fast and very good for data set which contains clusters of similar density \cite{scikit}. The most important input parameter is the maximum distance between two points for them to be considered as in the same neighborhood, $\epsilon$. We used $\epsilon = 10^{-2}$, as discussed in the main text, and the main output is the number of clusters. In our case, it reported 3 clusters, which is a very sounding estimation and fits well the Elbow curve for predicting the optimal $k$ for the $k$-means clustering. The Elbow method corresponds to a analysis designed to help finding the appropriate number of clusters in a data set. It plots the sum of the negative (Score method in scikit-learn KMeans package) of the clusters variance as a function of the assumed number of clusters. One should pick a number of clusters so that adding another cluster does not give much better modeling of the data set. In Fig. \ref{fig:elbow}, it is $k=3$.

\subsubsection{k-Nearest Neighbors}

The $k$-means algorithm has a vague relationship to the $K$-Nearest Neighbor (KNN) classifier \cite{altman} (not for the $k$ in the name, but because $k$-means classifier is very similar to KNN with $K=1$), so it was our first weapon when switching from unsupervised to supervised learning, given the success of the previous one. Roughly saying, the KNN method considers that similar objects are close to each other.

%Applying the 1-nearest neighbor classifier to the cluster centers obtained by k-means classifies new data into the existing clusters. This is known as nearest centroid classifier or Rocchio algorithm.

KNN is one of the simplest algorithms in ML using bassically the notion of distance in its construction. It is a neighbors-based classification where we perform instance-based learning or non-generalizing learning meaning that it does not construct a model, but simply stores instances of the training data. Classification is made from a simple majority vote of the nearest neighbors of each point: a query point is assigned the data class which has the most representatives within the nearest neighbors of the point.

The algorithm runs as follows: 1) set $K$ value;
2) For every instance $X$ (query example) in the data set: i) calculate the distance between the query example and true output, ii) add the distance and the corresponding index of the query example to an ordered collection; 3) Sort the ordered collection of distances and indices from smallest to largest (in ascending order) by the distances; 4) Pick the first $K$ entries from the sorted collection; 5) Get the labels of the selected $K$ entries; 6) Return the mode of the $K$ labels (classification tasks).

The code was made using the package \textit{sklearn.neighbors.KNeighborsClassifier} \cite{scikit}. The important parameter used was \textit{n\_neighbors}=7 ($K=7$). The default is the Euclidean distance, although it is possible to choose others such as Minkowsky or Manhattan distance. %This method is very fast and its computational cost is proportional to the number of points of the \textit{training set}.

\subsubsection{Random Forest} 

Random Forest \cite{breiman} is a powerful ensemble method widely used for hard and complex classification tasks. It consists of a multitude of Decision Trees. It outputs the most voted (mode) class of the individual decision trees. It is intuitive that training various models and taking their weighted predictions could be beneficial, once it mimics the ancient idea behind the "wisdom of the crowds". However, let us first define the Decision Tree learning. 

Decision Tree learning consists in the construction of a decision tree from labeled training instances. The root node is the training data itself, the branches nodes are the output (Boolean output) of tests for the attributes and leaves nodes are the estimated class. Therefore, for a given input $X_1$, a label $y_1$ is predicted after percolating from the root up to the leaf. In general, isolated, individual Decision Tree algorithm is a weak learner, producing high-bias error.

To implement a combination of Decision Trees, we first partition the training set in $M$ smaller subsets ${B_1, B_2, ..., B_M}$. The subsets are randomly chosen (and may be repeated). If the subsets are large enough for training a specific learner, they can be aggregated to create an ensemble predictor. For classification tasks, we take a majority vote of all the predictions. This process was introduced in Ref. \cite{breiman_dt} and is called $\textit{BAGGing}$, acronym for Bootstrap AGGregation. This can be demonstrated to reduce variance (out-of-sample error) without increasing the bias (in-sample error) \cite{highbias}.  In bagging, the contribution of the predictor is taken to be equal. 

The key idea of Random Forest is to perform subsets of the features. Now, the trees randomly choose a $k$ number, of $N$ total features with $k<N$. This bagging of features reduces the correlation between the various Decision Trees, contributing to better modelling. It can be formally demonstrated that a large amount of uncorrelated \textit{a priori} weak predictor can be aggragated to reduce variance  \cite{highbias}.

The code was made using the package \textit{sklearn.ensemble.RandomForestClassifier} \cite{scikit}. The important parameters used were \textit{max\_depth}=5, \textit{n\_estimators}=100.

\subsubsection{Extreme Gradient Boosting (XGB)}

Extreme Gradient Boosting (XGBoost) \cite{friedman} is another powerful ensemble technique. It is build upon the idea of the Boosting method. Unlike bagging, in boosting $k$ classifier contributes with different weights $\alpha_k$. So, we have here a weighted sum of the classifier predictors. It make sense, once one can imagine that, for instance, the "opinion" of a better classifier (measured on the test set) should contribute more in comparison to weaker classifier, which is sometimes termed as "autocratic" approach in contrast with the "democratic" view in the bagging method.

In XGB, the idea is even more sophisticated. Here, the addition of new Decision Trees is done if it contributes to minimizing a cost function of the ensemble. In this manner, XGB provides a clever way to map the gradient to Decision Trees. The technical details are far beyond the scope of this Appendix, but the implication by means of the Scikit-Learn API \textit{XGBClassifier} is trivial.  The main parameters used were \textit{max\_depth}=5, \textit{n\_estimators}=100.

\subsubsection{Multilayer Perceptron (MLP)} 

Multilayer Perceptrons are the pillar of what is termed as deep learning modeling. It is a sophisticated supervised learning technique in the class of artificial feed-forward neural networks (or simply neural nets) used to approximate an unknown function $f(\mathbf{X})$ by $f^{*}(\mathbf{X};\boldsymbol{\theta})$, which maps an input $\mathbf{X}$ to an output $\mathbf{y}$. Once the learning process is concluded back, it retrieves the optimal $\boldsymbol{\theta}$ \cite{highbias, rosenblatt}. In general, the larger the dimension of the vector $\boldsymbol{\theta}$ (the number of parameters), the better the approximation $f^{*}$ for complex tasks can be.
 
\begin{figure}[t] \centering
\includegraphics*[scale =1]{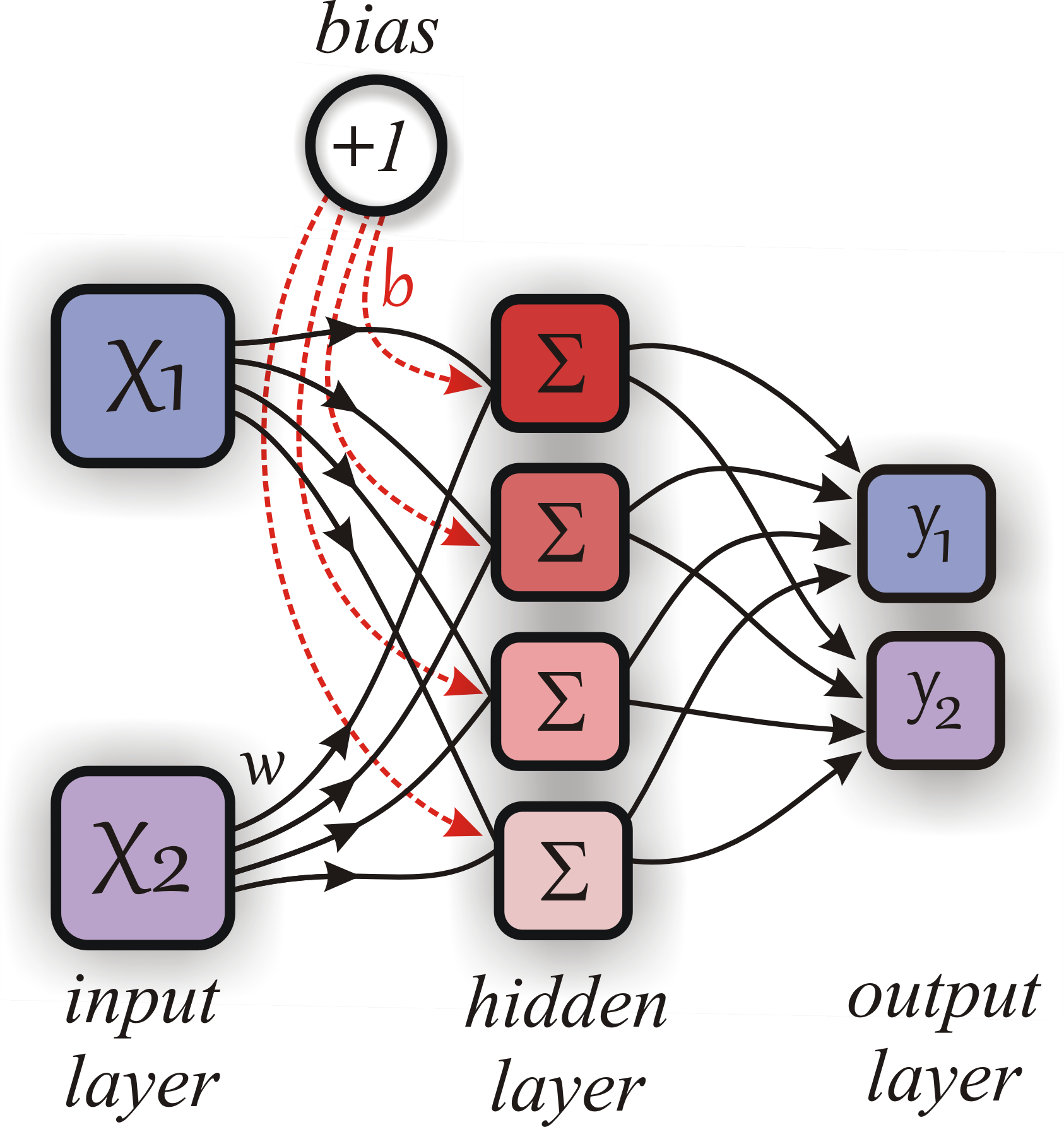}
\caption{Architecture of an artificial neural network composed of three layers with one hidden layer. The nodes stand for neurons outputs and the solid (dashed) arrows represent neuron-specific weight (bias). Each neuron processes the incoming signals by using the same activation function $\Sigma$.}
\label{f27}
\end{figure}

They were initially formulated to emulate a nervous system (the reason for the term ``neural'') composed of a multitude of basic units or neurons stacked into layers, in which the output of one layer is the input for the succeeding layer. They are said to be feed-forward in virtue of the one-way flow from the input layer to the output layer, see Fig. \ref{f27}.

The connections between layers are performed by means of the combination of the inputs provided by all neurons in the previous layers (together with a re-centering bias) and all neurons in current layer $i$ (see solid and dashed arrows in Fig. \ref{f27}) given by 
\begin{eqnarray}
\mathbf{z}_{i}=\mathbf{w}_{i}\cdot \mathbf{x} + \mathbf{b}_{i},
\end{eqnarray}
where $\mathbf{x}$ is the input vector (output of previous layer), $\mathbf{w}_{i}$ and $b_i$ are the vectors of neuron-specific weights and bias for the $i$ layer, respectively \cite{highbias}. The bias acts similarly to the non-null intercept in a linear regression problem, augmenting the space of tentative solutions, enhancing the power of representation (or expressivity) of the model \cite{highbias}.

Paradoxically, for first layer $\mathbf{z}_{i} = \mathbf{X}$, i. e., the input layer just outputs the features themselves to the next layer. For the hidden layer, each neuron $j$ outputs a scalar after processing a non-linear transformation of inputs it previously received. For instance, if the activation function is a (Heaviside) step function, a given neuron receives various signals from the other neurons in the previous layer and "decides" if it should activate (add up of the signals is greater than zero, outputting 1) or not (add up of the signals is less than zero, outputting 0). This step function was the original nonlinearity implemented in the MLP, hence the denomination "activation function". Currently, usual activation functions are, for instance, the logistic function ($\Sigma(x)=1/(1+\exp^{-x})$) and the hyperbolic tangent ($\Sigma(x)=\tanh{x}$). The number of hidden layers and activate function are hyperarameters that one can adjust to achieve better predictions. In this work, we used the rectified linear unit (ReLU) function, $\Sigma(x) = \text{max}(0, x)$. Not only because of increasing accuracy, but also due to the reduction in the computational time, once it has be shown to be optimal in computing the gradient which updates the parameter of the model $\mathbf{w}$ and $\mathbf{b}$ via the back-propagation method \cite{rumelhart}. This is easily done and requires one to set the 'activation = relu' in the package \textit{sklearn.neural$\_$network.MLPClassifier} \cite{scikit}. Moreover, we used two hidden layers of 100 neurons each in our research. The middle layers are "hidden" layers because it is not known what they should output for the next layer in order to reach the aim of finding a value $y*$ close to $y$ for each $X$ in $\mathbf{X}$. 
When the output layer is reached, it performs, in general, a simple logistic regression or soft-max for classification tasks or linear regression for regression problems.  

Given a cost function $J(\boldsymbol{\theta})$, for instance 

\begin{eqnarray}
J(\boldsymbol{\theta}) = \frac{1}{m} \sum_{\mathbf{x} \in \mathbf{X}^{\text{train}} } (\mathbf{y}^{\text{train}} - f^{*}(\mathbf{x};\boldsymbol{\theta}) )^2,
\end{eqnarray}
where $m$ represents the training set size, we encounter the hard problem of computing the gradient, $\nabla_{\boldsymbol{\theta}} J(\boldsymbol{\theta})$, of a high dimensional and complex function. Even for a model with a few hidden layers of hundreds of neurons we have thousands or millions of parameters. The best $\boldsymbol{\theta}$ returns the values of the weights $\bf{w}$ (solid arrows in Fig. \ref{f27}) and the bias $\bf{b}$ (dashed arrows in in Fig. \ref{f27}) which give the best generalization.  One overcomes this computational cost by minimizing the error in the direction from the output layer to the input layer with the back-propagation algorithm and its variants. In our codes, we used the \textit{Adam} algorithm by just setting "solver = adam" in the scikit-learn MLP package, see Ref. \cite{adam} for details.

%Each layer $i$ can be considered a function $f^{(i)}$ and the model is therefore a network or composition of functions, $i$ representing the depth of the deep learning machine. They are also fully connected as every neuron of a given layer is connected to every neuron of the next layer.

\end{document}